\documentclass[pdftex,twocolumn,epjc3]{svjour3}          

\RequirePackage[T1]{fontenc}

\smartqed  

\RequirePackage{graphicx}
\RequirePackage{mathptmx}      
\RequirePackage{flushend}
\RequirePackage[numbers,sort&compress]{natbib}
\RequirePackage[colorlinks,citecolor=blue,urlcolor=blue,linkcolor=blue]{hyperref}

\usepackage{graphicx}
\usepackage{epsfig}
\usepackage{latexsym}
\usepackage{dialogue}
\usepackage{times}
\usepackage{url}
\usepackage{wrapfig}
\usepackage{fullpage}
\usepackage{caption}
\usepackage{subcaption}

\usepackage{enumitem} 

\journalname{To appear in ``Ethics of Information Technology'' -- }

\begin{document}

\title{Society-in-the-Loop: Programming the Algorithmic Social Contract}

\author{Iyad Rahwan\thanksref{addr1,addr2,t1}}


\thankstext[$\star$]{t1}{An earlier version of this article was published under the same title on \url{medium.com}, on August 12, 2016.}
\thankstext{e1}{e-mail: irahwan@mit.edu}

\institute{The Media Lab, Massachusetts Institute of Technology, Cambridge MA, 02139 USA\label{addr1}
       \and
     Institute for Data, Systems and Society, Massachusetts Institute of Technology, Cambridge MA, 02139 USA\label{addr2}
}

\date{This version: 20 July 2017}

\maketitle

\begin{abstract}
Recent rapid advances in Artificial Intelligence (AI) and Machine Learning have raised many questions about the regulatory and governance mechanisms for autonomous machines. Many commentators,  scholars, and policy-makers now call for ensuring that algorithms governing our lives are transparent, fair, and accountable. Here, I propose a conceptual framework for the regulation of AI and algorithmic systems. I argue that we need tools to program, debug and maintain an \emph{algorithmic social contract}, a pact between various human stakeholders, mediated by machines. To achieve this, we can adapt the concept of \emph{human-in-the-loop} (HITL) from the fields of modeling and simulation, and interactive machine learning. In particular, I propose an agenda I call \emph{society-in-the-loop} (SITL), which combines the HITL control paradigm with mechanisms for negotiating the values of various stakeholders affected by AI systems, and monitoring compliance with the agreement. In short, `\emph{SITL} = \emph{HITL} + \emph{Social Contract}.'
\end{abstract}

\section{Introduction}

\begin{quote}
\textit{``Art goes yet further, imitating that Rationall and most excellent worke of Nature, Man. For by Art is created that great LEVIATHAN called a COMMON-WEALTH, or STATE, (in latine CIVITAS) which is but an Artificiall Man''}\\ 
Thomas Hobbes (1651). Leviathan
\end{quote}



Despite the initial promise of Artificial Intelligence, a long `AI Winter' ensued in the 1980s and 1990s, as problems of automated reasoning proved much harder than initially anticipated \cite{markoff2015machines}. But recent years have seen rapid theoretical and practical advances in many areas of AI. Prominent examples include machines learning their own representations of the world via Deep Neural Network architectures \cite{lecun2015deep}, Reinforcement Learning from evaluative feedback \cite{littman2015reinforcement}, and economic reasoning in markets and other multi-agent systems \cite{parkes2015economic}. The result is an accelerating proliferation of AI technologies in everyday life \cite{levy2010ai}.

These advances are yielding substantial societal benefits, ranging from more efficient supply chain management, to better matchmaking in peer-to-peer markets and online dating apps, to more reliable medical diagnosis and drug discovery \cite{100years}. 

But AI advances have also raised many questions about the regulatory and governance mechanisms for autonomous machines and complex algorithmic systems. Some commentators are concerned that algorithmic systems are not accountable because they are \emph{black boxes} whose inner workings are not transparent to all stakeholders \cite{pasquale2015black}. Others raised concern over people unwittingly living in filter bubbles created by news recommendation algorithms \cite{pariser2011filter,bozdag2013bias}. Others argue that data-driven decision-support systems can perpetuate injustice, because they can also be biased either in their design, or by picking up human biases in their training data \cite{tufekci2015algorithmic,caliskan2017semantics}. Furthermore, algorithms can create feedback loops that reinforce inequality \cite{boyd2012critical}, for example in the use of AI in \emph{predictive policing} or \emph{creditworthiness} prediction, making it difficult for individuals to escape the vicious cycle of poverty \cite{o2016weapons}.

In response to these alarms, various academic and governmental entities have started thinking seriously about AI governance. Recently, the United States White House National Science and Technology Council Committee on Technology released a report with recommendations ranging from eliminating bias from data, to regulating autonomous vehicles, to introducing ethical training to computer science curricula \cite{whitehouse}. The European Union, which has enacted many personal data privacy regulations, will soon vote on a proposal to grant robots legal status in order to hold them accountable, and to produce a code of ethical conduct for their design \cite{EU}. The Institute of Electrical and Electronics Engineers recently published a vision on `Ethically Aligned Design' \cite{ieee}. Industry leaders have also taken the initiative to create a `Partnership on AI' to establish best practices for AI systems and to educate the public about AI \cite{partnership}.

My goal is this paper is to introduce a conceptual framework for thinking about the regulation of AI and data-driven systems.  I argue that we need a new kind of social contract: an \emph{algorithmic social contract}, that is a contract between various stakeholders, mediated by machines. To achieve this, we need to adopt a \emph{society-in-the-loop} (SITL) framework in thinking about AI systems, which adapts the concept of \emph{human-in-the-loop} (HITL), from the fields of supervisory control and interactive machine learning, but extends it to oversight conducted by society as a whole.

\section{Human-in-the-Loop}

\begin{figure}[h]
  \centering
	\includegraphics[width=0.48\textwidth]{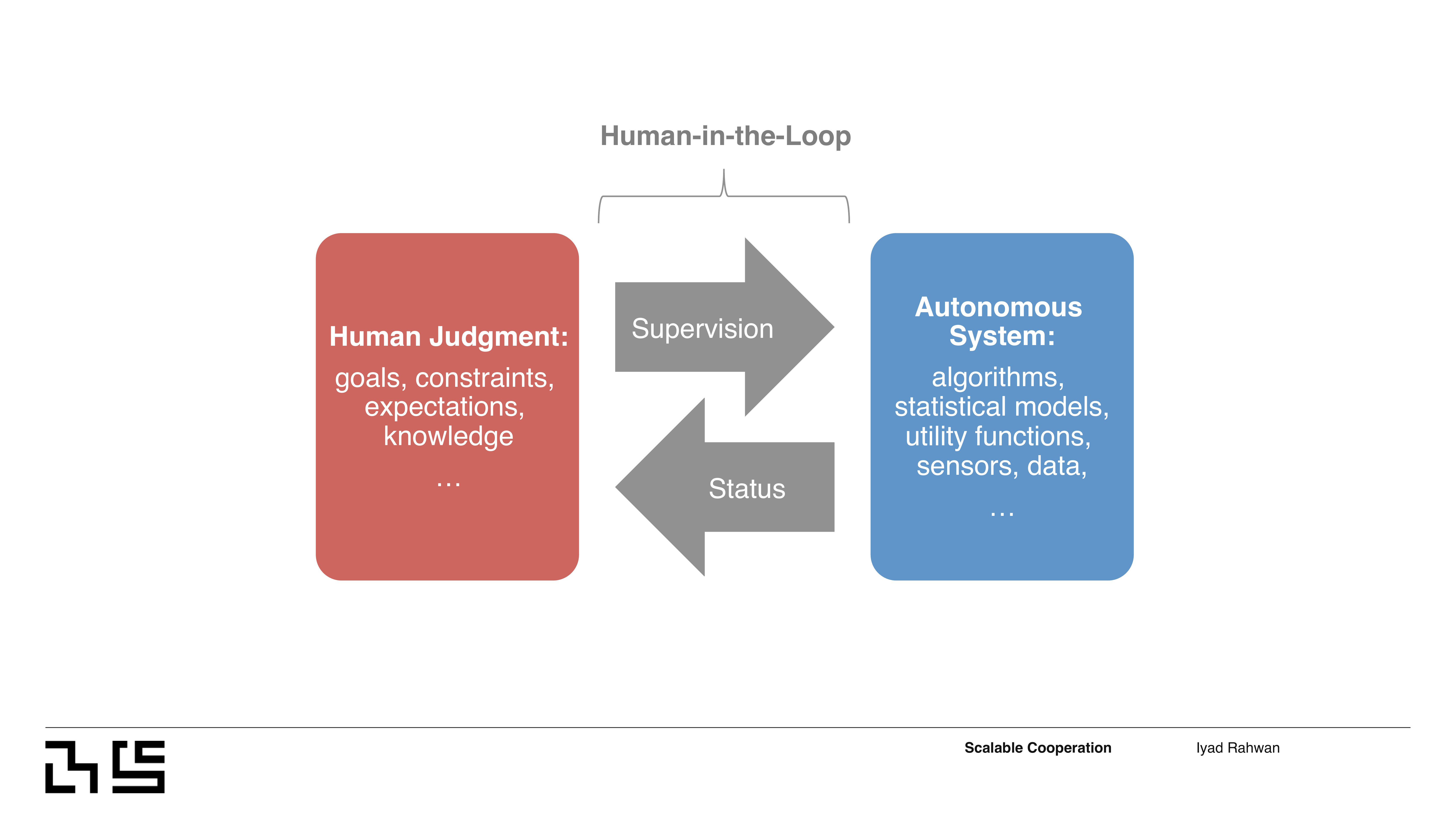} 
	\caption{In a HITL system, a human provides monitoring and supervisory functions at crucial junctions in the system's operation.}
  \label{fig:HITL}
\end{figure}

In a \emph{human-in-the-loop} (HITL) system,  a human operator is a crucial component of an automated control process, handling challenging tasks of supervision, exception control, optimization and maintenance (Figure \ref{fig:HITL}). The notion has been studied for decades within the field of \emph{supervisory control} \cite{sheridan2006supervisory,allen1999mixed}. Sheridan defined \emph{human supervisory control} as a process by which ``one or more human operators are intermittently programming and continually receiving information from a computer that itself closes an autonomous control loop through artificial effectors to the controlled process or task environment'' \cite{sheridan1992telerobotics}.

These ideas then made their way into the field of Human-Computer Interaction (HCI). Scientists began working on \emph{mixed-initiative} user interfaces, in which the autonomous system can make intelligent decisions about when and how to engage the human \cite{horvitz1999principles}.


Recently, a number of articles have been written about the importance of applying HITL thinking to Artificial Intelligence (AI) and machine learning (ML) systems. A simple form of HITL ML is the use of human workers to label data for training machine learning algorithms. This has produced invaluable benchmarks that spurred major advances in computer vision, for example \cite{russakovsky2015imagenet}. 

Another example of HITL ML is \emph{interactive machine learning}, which can help machines learn faster or more effectively by integrating feedback interactively from users \cite{oreilly,amershi2014power}. This type of HITL ML has been going on for a while. For example, many computer applications learn from your behavior in order to improve their ability to serve you better (e.g. by predicting the next word you are going to type). Similarly, when you mark an email as `spam' in an online email service, you are one of many humans in the loop of a complex machine learning algorithm (specifically an active learning system), helping it in its continuous quest to improve email classification as spam or non-spam.

More sophisticated examples of HITL ML are now emerging, in which the human-in-the-loop has more explicit knowledge of the state of the system. For instance, in a crisis counseling system, a machine learning system classifies messages sent by callers, and provides visualizations to a human counselor in real-time \cite{dinakar2015mixed}. Thus, the human and the machine learning system work in tandem to deliver effective counseling.

HITL thinking has also been applied successfully to human-robot interaction (HRI) \cite{cakmak2010designing}. This includes dynamically adapting the degree of autonomy given to robots \cite{crandall2001experiments,tambe2002adjustable}, interactively teaching reinforcement learning robots to adopt particular behaviors \cite{thomaz2008teachable}, and designing flexible human-robot teams  \cite{johnson2014coactive}. 

There is another role of the HITL paradigm, which is closer to the problems discussed in the present article. HITL is not only a means to \emph{improve} AI systems' accuracy in classification or to speed up the convergence of a reinforcement learning robot. Rather, HITL can also be a powerful tool for \emph{regulating} the behavior of AI systems. For instance, many scholars now advocate for expert oversight, by a human operator, over the behavior of `killer robots' or credit scoring algorithms \cite{citron2014scored}.

The presence of a human fulfills two major functions in a HITL AI system: 

\begin{enumerate}
\item The human can \emph{identify misbehavior} by an otherwise autonomous system, and take corrective action. For instance, a credit scoring system may mis-classify an adult as ineligible for credit, due to an error in data entry in their age--something a human may spot from the applicant's photograph. Similarly, a computer vision system on a weaponized drone may mis-identify a civilian as a combatant, and the human operator--it is hoped--would ensure that such cases are identified, and override the system. Some work is underway to ensure AI cannot learn to disable their own kill-switch \cite{safely}.

\item The human can be involved in order to provide an \emph{accountable entity} in case the system misbehaves. If a fully autonomous system causes harm to human beings, having a human in the loop provides trust that somebody would bare the consequence of such mistakes, and thus have incentive to minimize those mistakes. This person may be a human within a tight control loop (e.g. an operator of a drone) or a much slower loop (e.g. programmers in a multi-year development cycle of an autonomous vehicle). Until we find a way to punish algorithms for harm to humans, it is hard to think of any other alternative.
\end{enumerate}

While HITL is a useful interaction paradigm for building AI systems that are subject to oversight, I believe it does not sufficiently emphasize the role of society as a whole in such oversight. HITL suggests that once we put a human expert, or group of experts, within the loop of an AI system, the problem of regulation is solved. But as I shall discuss in the following section, this may not always be the case.


\section{Society-in-the-Loop}

What happens when an AI system does not serve a narrow, well-defined function, but a broad function with wide societal implications? Consider an AI algorithm that controls millions of self-driving cars; or a set of news filtering algorithms that influence the political beliefs and preferences of millions of citizens; or algorithms that mediate the allocation of resources and labor in an entire economy. What is the HITL equivalent of these algorithms? This is where we make the qualitative shift from HITL to \emph{society in the loop} (SITL).

While HITL AI is about embedding the judgment of \emph{individual} humans or groups in the optimization of AI systems with \emph{narrow impact}, SITL is about embedding the values of \emph{society}, as a whole, in the algorithmic governance of societal outcomes that have \emph{broad implications}. In other words, SITL becomes relevant when the scope of both the input and the output of AI systems is very broad. But one might ask, why should this be any different?

The move from HITL to SITL raises a fundamentally different problem: how to balance the competing interests of different stakeholders, including the interests of those who govern through algorithms? This is, traditionally, a problem of defining a \emph{social contract} \cite{skyrms2014evolution}. To put it in the most skeletal form, we can say:

\begin{quote}
\begin{center}
\fbox{\begin{minipage}{14em}
SITL = HITL + Social Contract
\end{minipage}}
\end{center}
\end{quote}

To elaborate on this simple equation, we need to take a short detour into political philosophy. 

\section{Detour: The Social Contract}

Humans are the ultimate cooperative species \cite{nowak2011supercooperators}. Cultural anthropologists trace the evolution of political systems of governance from decentralized bands and tribes, to increasingly centralized chiefdoms, sovereign states and empires \cite{haviland2013cultural}. 

Over time, humans reached the limits of old cooperative institutions such as \emph{kin selection}--helping others who share their genes \cite{hamilton1963evolution}, and \emph{reciprocal altruism}--helping others who would later help them back \cite{trivers1971evolution}. These old mechanisms cannot scale adequately to larger groups. In the face of inter-group competition, evolutionary pressure favored the emergence, and spread, of more complex social institutions to coordinate people's behaviors \cite{turchin2015ultrasociety,young2001individual}. For example, centralized sanctioning power is able to prevent higher-order free-riding--following cooperative norms, but not contributing to their enforcement--that undermine cooperation in larger groups \cite{gurerk2006competitive,baldassarri2011centralized,sigmund2010social}. 

\begin{figure*}[t]
  \centering
	\includegraphics[width=0.9\textwidth]{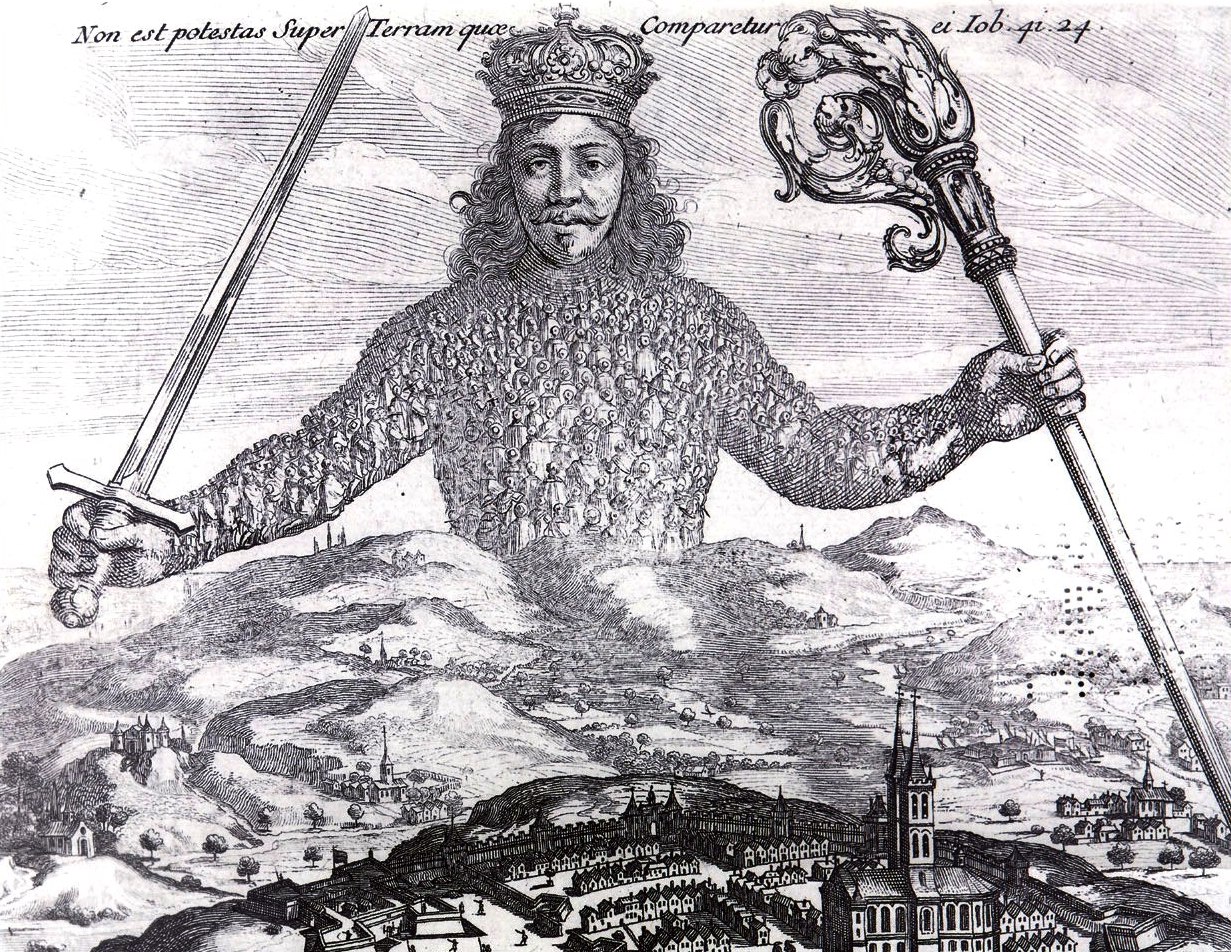} 
	\caption{The frontispiece of Thomas Hobbes' 1651 book \emph{Leviathan} by Parisian artist Abraham Bosse. The piece is a striking depiction of how the sovereign--a giant ruling a peaceful realm through the warrior's sword and the monk's crosier. The torso and arms of the figure are composed of over three hundred persons, signifying that the Leviathan derives his power to govern, not from divine authority, but through the consent of the governed.}
  \label{fig:leviathan}
\end{figure*}

The founders of \emph{social contract} theory, going back to Thomas Hobbes' landmark book, \emph{Leviathan} \cite{leviathan}, posit that centralized government is legitimate precisely because it enables industrious people to cooperate via third-party enforcement of contracts among strangers (see Figure \ref{fig:leviathan}). Some of these contracts are explicit, such as marriage contracts or commercial transactions. Other aspects of the social contract are implicit, being embedded in social norms that govern every day life. In both cases, the contract embodies mutual consent to the government's legitimate use of force--or people's use of social pressure-- to guard people's rights and punish violators \cite{skyrms2014evolution,binmore2005natural}.

Hobbes gave his Leviathan, the sovereign, enormous power. Subsequently, the social contract undertook many stages of evolution, thanks to enlightenment thinkers like John Locke \cite{locke1689two}, Jean-Jacques Rousseau \cite{rousseau1762social}, all the way to  John Rawls \cite{rawls1971theory} David Gauthier \cite{gauthier1986morals} and Brian Skyrms \cite{skyrms2014evolution} in modern times. These thinkers refined our conception of how the social contract emerges in the first place, as well as the ways in which we can keep it from collapsing.

Modern political institutions, including the modern state, are a product of these evolutionary mechanisms of political development, which combine institutional innovation with learning. As Fukuyama puts it, ``[s]ocieties are not trapped by their pasts and freely borrow ideas and institutions from each other'' \cite{fukuyama2011origins}. 

The result of this evolutionary process is a social contract that can provide the efficiency and stability of sovereign states, but which also ensures the sovereign implements the \emph{general will} \cite{rousseau1762social} of the people, and is held in some way accountable for violations of fundamental rights. In the same manner, ``[n]ew algorithmic decisionmakers are sovereign over important aspects of individual lives'' Thus, lack of accountability and due process for algorithmic decisions risks ``paving the way to a new feudal order'' \cite{citron2014scored}.

\begin{figure*}[t]
  \centering
	\includegraphics[width=0.9\textwidth]{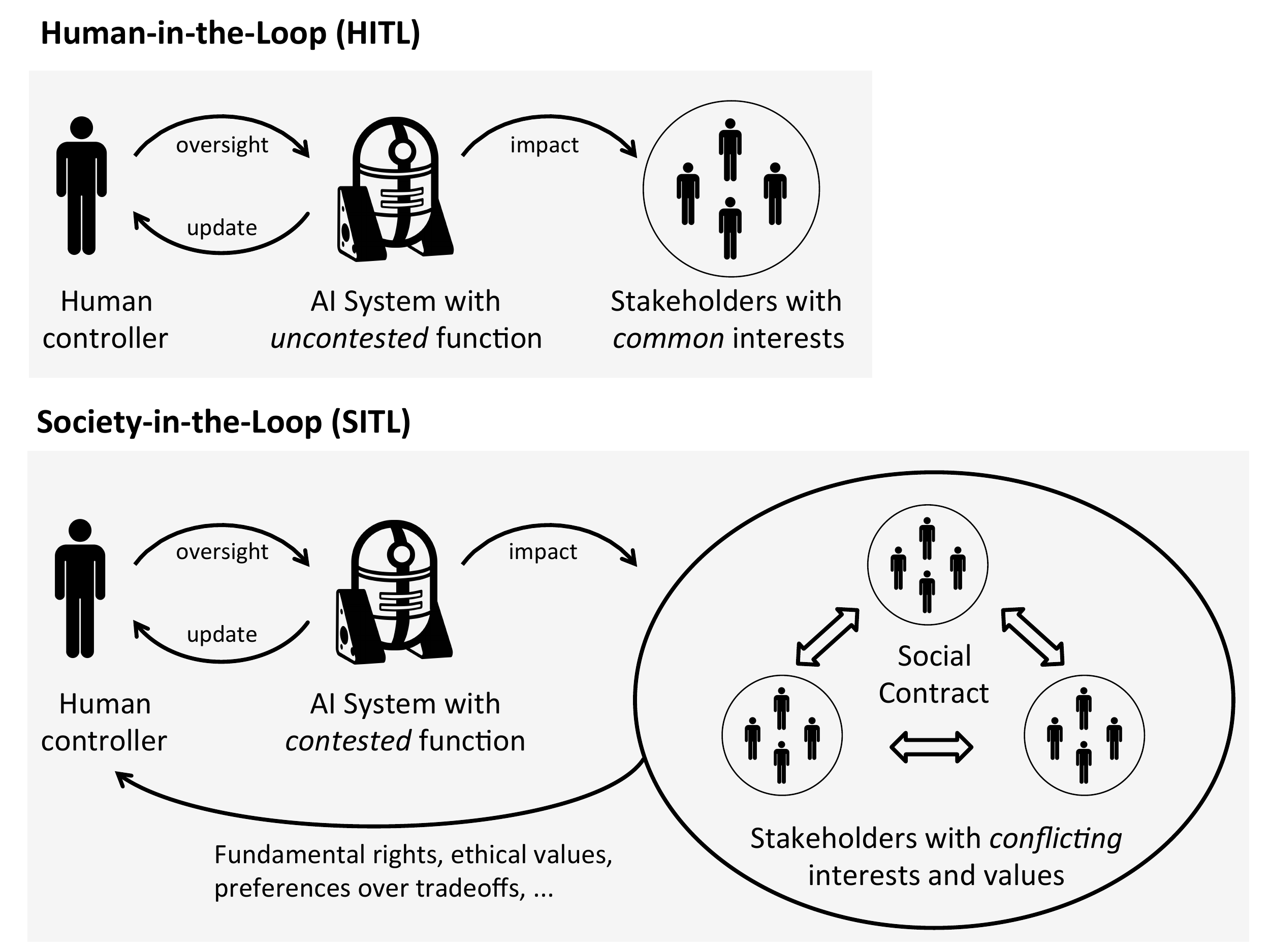} 
	\caption{Society-in-the-Loop (SITL) = Human-in-the-Loop (HITL) + Social Contract; \emph{(Top)} In a HITL system, a human controller monitors and exercises oversight over the operation of an AI system to ensure that it serves the uncontested and common goals of its stakeholders. For example, a human pilot oversees an airplane autopilot to increase passenger safety. \emph{(Bottom)} In SITL, the AI system has broad impact, requiring various societal stakeholders to identify the fundamental rights that the AI must respect, the ethical values that should guide the AI's operation, the cost and benefit tradeoffs the AI can make between various stakeholder groups, etc.}
  \label{fig:HITL_vs_SITL}
\end{figure*}

\section{The Algorithmic Social Contract}

The SITL paradigm that I advocate is more akin to the interaction between a government and a governed citizenry, than the interaction between a drone and its operator. Similar to the role of due process and accountability in the traditional social contract, SITL can be conceived as an attempt to embed the general will into an \emph{algorithmic} social contract.

By using the social contract metaphor, the SITL paradigm emphasizes an important distinction from the traditional HITL paradigm (Figure \ref{fig:HITL_vs_SITL}). In a HITL system, a human controller ensures that the AI system fulfills \emph{uncontested} and \emph{common} goals on behalf of societal stakeholders--e.g. ensuring a plane lands safely, or improving food quality inspection. In addition, in the SITL domain, society must agree on two aspects:
\begin{enumerate}
\item Society must resolve tradeoffs between the different values that AI systems can strive towards--e.g. tradeoffs between security and privacy, or the tradeoffs between different notions of fairness \cite{berk2017fairness,kleinberg2016inherent}.
\item Society must agree on which stakeholders would reap which benefits and pay which costs--e.g. how improvements in safety made possible by driverless cars are to be distributed between passengers and pedestrians, or which degree of collateral damage, if any, is acceptable in autonomous warfare.
\end{enumerate}

In human-based government, citizens use various channels---e.g. democratic voting, opinion polls, civil society institutions, the media---to articulate their expectations to the government. Meanwhile, the government, through its bureaucracy and various branches undertakes the function of governing, and is ultimately evaluated by the citizenry. And while citizens are not involved in the details \cite{lippmann1927phantom}, they are the arbiters among all of these institutions, and have the power to replace their key actors.

Modern societies are (in theory) SITL human-based governance machines. Some of those machines are better programmed, and have better `user interfaces' than others. Similarly, as more governance functions get encoded into AI algorithms, we need to create channels between human values and governance algorithms.

To implement SITL, we need to know what types of behaviors people expect from AI, and to enable policy-makers and the public to articulate these expectations (goals, ethics, norms, social contract) to machines. To close the loop, we also need new metrics and methods to evaluate AI behavior against quantifiable human values. In other words: we need to build new tools to program, debug, and monitor the algorithmic social contract between humans and algorithms--that is, algorithms that are effective sovereigns over important aspects of social and economic life, whether or not they are actually operated by governments. This requires both government regulation and industry standards that represent the expectations of the public, with corresponding oversight.

\section{The SITL Gap}

Why are we not there yet? There has been a flurry of thoughtful treaties on the social and legal challenges posed by the opaque algorithms that permeate and govern our lives. While these seminal writings help illuminate many of the challenges, they fall short on comprehensive solutions. 

\subsection{Articulating Societal Values}

\begin{figure}[h]
  \centering
	\includegraphics[width=0.48\textwidth]{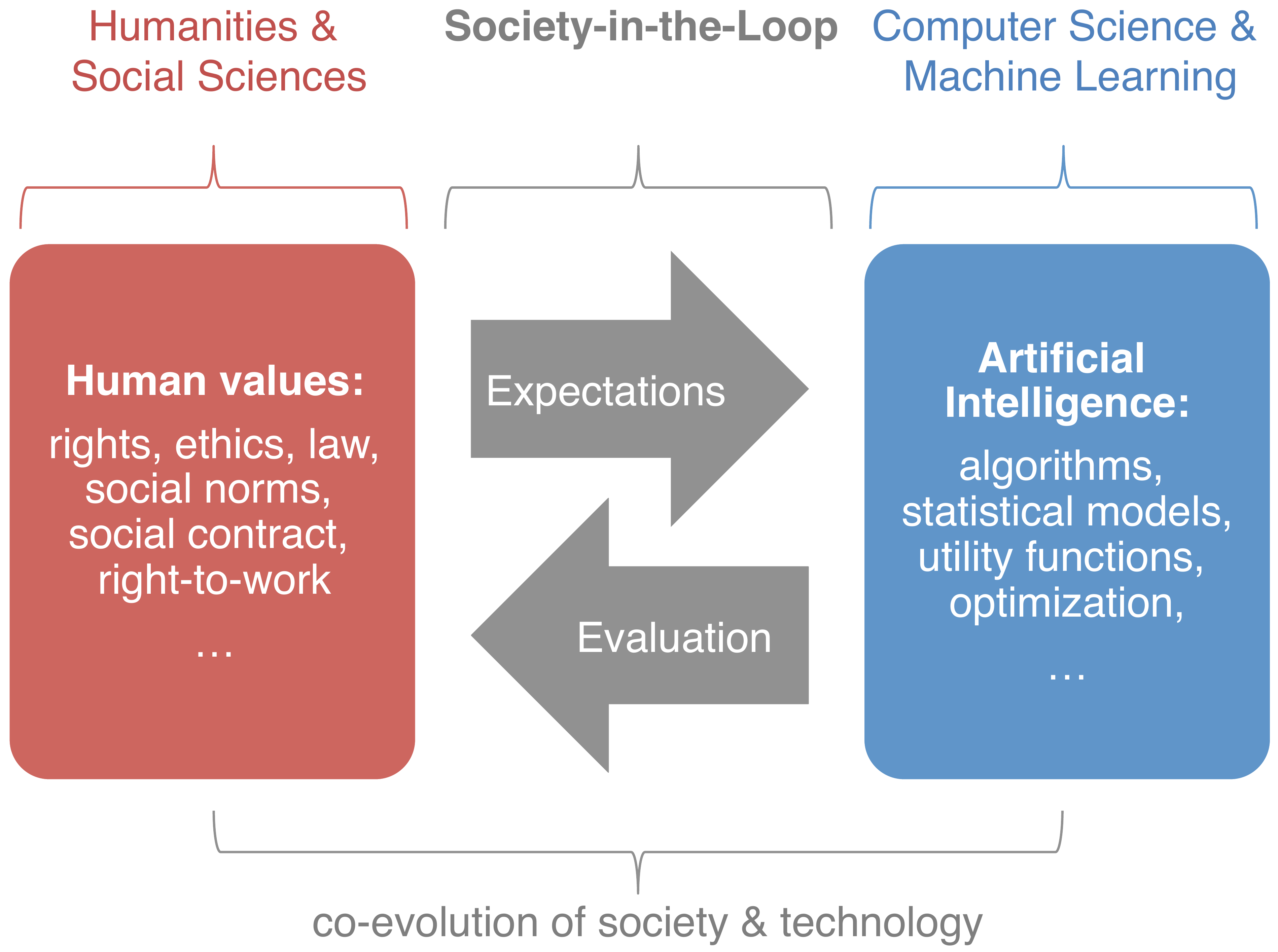} 
	\caption{In a SITL system, broad societal values (as opposed to an individual human operator's judgment) must be involved in the monitoring and supervisory function of AI systems that have wide-ranging societal implications (as opposed to AI systems with narrow impact).}
  \label{fig:SITL2}
\end{figure}

One barrier to implementing SITL is the cultural divide between engineering on one hand, and the humanities on the other (see Figure \ref{fig:SITL2}). Thoughtful legislators, legal scholars, media theorists, and ethicists are very skilled at revealing moral hazards, and identifying ways in which moral principles and constitutional rights may be violated \cite{castelfranchi2000artificial}. But they are not always able to articulate those expectations in ways that engineers and designers can operationalize.

\subsection{Quantifying Externalities \& Negotiating Tradeoffs}

Algorithms can generate what economists refer to as \emph{negative externalities}---costs incurred by third parties not involved in the decision \cite{pigou1920}. For example, if autonomous vehicle algorithms over-prioritize the safety of passengers---who own them or pay to use them---they may disproportionately increase the risk borne by pedestrians. Quantifying these kinds of externalities is not always straightforward, especially when they occur as a consequence of long, indirect causal chains, or as a result of machine code that is opaque to humans.

Once we have quantified externalities, we need to negotiate the \emph{tradeoffs} they embody. If certain ways to increase pedestrian safety in autonomous vehicles imply reduction in passenger safety, which tradeoffs are acceptable?

Human experts already implement tradeoffs as they design policies and products. For example, reducing the speed limit on a road reduces the utility for drivers who want to get home quickly, while increasing the overall safety of drivers and pedestrians. It is possible to completely eliminate accidents---by reducing the speed limit to zero and banning cars---but this would also eliminate the utility of driving, and regulators attempt to strike a balance that society is comfortable with through a constant learning process.

Quantifying tradeoffs in any complex system, with many interacting parts, is always difficult. In complex economic systems, there are often \emph{unintended consequences} of design choices. As AI becomes an integral part of such systems, the problem of quantifying those tradeoffs becomes even harder. For example, subtle algorithm design choices in autonomous vehicles may lead to a particular tradeoff between risks to passengers and risks to pedestrians. Identifying, let alone negotiating those tradeoffs, may be much harder than setting a speed limit--if only due to the greater degrees of freedom when making design choices. This may be further complicated by the fact that algorithms learn from their experience, which may lead to shifts in the tradeoffs being made, going beyond what the programmers intended.

\subsection{Verifying Compliance with Societal Values}

Computer scientists and engineers are not always able to quantify the behaviors of their systems such that they can be easily understood by ethicists and legal theorists. This makes it more difficult to scrutinize the behavior of algorithms against set expectations. Even simple notions such as `fairness' can be formalized in many different ways mathematically or in computer code \cite{berk2017fairness}.

An important component of Figure \ref{fig:SITL2} is that both human values and AI are ongoing constant co-evolution. Thus, the evolution of technical capability can dramatically (and even irreversibly) alter what society considers acceptable---think of how privacy norms have changed because of the utility provided by smart phones and the Internet.

\section{Bridging the Gap}

There are many efforts underway to bridge the society-in-the-loop gap. Below is an incomplete list of efforts that I believe are relevant, and a discussion of their merits and limitations.

\subsection{Articulating Values: Design, Crowdsourcing \& Sentiment Analysis}

In the broader context of technology design, various \emph{value-sensitive design} methodologies have been proposed \cite{friedman1996value}, which can be applied to software development \cite{aldewereld2014design,van2013translating}. These approaches may prove helpful in the design of AI systems.

Some AI scientists propose to use of crowdsourcing \cite{conitzer2015crowdsourcing} to identify societal tradeoffs in a programmable way. There are some efforts to collect data about people's preferences over values implemented in AI algorithms, such as those that control driverless cars. Using methods from the field of moral psychology, one can identify potential moral hazards due to the incentives of different users of the road \cite{socialdilemma}. For example, my co-authors and I have developed a public-facing survey tool that elicits the public's moral expectations from autonomous cars faced with ethical dilemmas \cite{MoralMachine}. We have collected over 30 million decisions to date. Findings from this data can help regulators and car makers understand some of the psychological barriers to the wide adoption of autonomous vehicles.

In many domains, it may be possible to measure societal values directly from observational data, without having to run explicit polling campaigns or build dedicated crowdsourcing platforms \cite{liu2012sentiment}. For example, automated sentiment analysis on social media discourse can quantify people's reaction to different moral violations committed by AI systems. While these approaches have their limitations, they can help gauge the evolution of public attitudes, and their readiness to accept new social pacts through machines.

\subsection{Negotiation: Social Choice \& Contractarianism}

The field of \emph{computational social choice} \cite{moulin2016handbook,arrow2012social} explores the aggregation of societal preferences and fair allocation of resources. Because these aggregation mechanisms can be implemented algorithmically, they provide a potential solution to the problem of negotiating tradeoffs of different stakeholders \cite{nisan2007algorithmic,chen2013truth,parkes2015economic}.

An alternative approach to the negotiation of values is to use normative and meta-ethical tools from social contract theory to identify enforceable outcomes that rational actors would be willing to opt into. For instance, Leben recently proposed an algorithm that allows autonomous vehicles to resolve dilemmas of unavoidable harm using Rawls' Contractarianism \cite{leben2017rawlsian}. In particular, Leben proposes to program cars to make decisions that rational actors would take if they were in a hypothetical `original position' behind a  `veil of ignorance.' This veil would, for example, conceal whether the person is a passenger or a pedestrian in a given accident, leading them to choose the \emph{maximin} solution--that is, the decision that minimizes how bad the worse-case outcome is.

\subsection{Compliance: People Watching Algorithms}

An important function for ensuring accountability is the ability to scrutinize the behavior of those in power, through mechanisms of trasparency. In the context of algorithms, this does \emph{not} mean having access to computer source code, as intuitive as this notion might seem.

Reading the source code of a modern machine learning algorithm tells us little about its behavior, because it is often through the interaction between algorithms and data that things like discrimination emerge. Transparency must, therefore, be about the external behavior of algorithms. Indeed, this is how we regulate the behavior of humans--not by looking into their brain's neural circuitry, but by observing their behavior and judging it against certain standards of conduct. Of course, this observation can benefit from the ability of the algorithm to give human-interpretable explanations of their decisions \cite{letham2015interpretable}.

The new journalistic practice of \emph{algorithmic accountability reporting} provides a framework for  scrutiny of algorithmic decisions that is purely behavioral \cite{diakopoulos2015algorithmic}. As an example, Sweeney has demonstrated that Web searches for names common among African Americans cause online advertising algorithms to serve ads suggestive of an arrest record, which can harm the individual being searched \cite{sweeney2013discrimination}. Investigative journalism has also revealed evidence of price discrimination based on users' information, sparking a debate about the appropriateness of this practice \cite{price}.

We might also envision a role for professional \emph{algorithm auditors}, people who interrogate algorithms to ensure compliance with pre-set standards. This interrogation may utilize real or synthetic datasets designed to identify whether an algorithm violates certain requirements. For instance, an algorithm auditor may provide sample job applications to identify if a job matching algorithm is discriminating between candidates based on irrelevant factors. Or an autonomous vehicle algorithm auditor may provide simulated traffic scenarios to ensure the vehicle is not disproportionately increasing the risk to pedestrians or cyclists in favor of passengers.

One weakness of auditing in a simulated environment--using computer simulation or fake data--is the potential for adversarial behavior: the algorithm being audited may attempt to trick the algorithm doing the auditing. This is similar to `defeat devices,' a term used to describe software or hardware features that interfere with or disables car emissions controls under real world driving conditions, even if the vehicle passes formal emissions testing \cite{defeat}. In a similar fashion, an autonomous vehicle control algorithm may detect that it is being tested in a virtual environment--e.g. by noticing that the distribution of scenarios is skewed towards ethical dilemmas--and behave differently under such testing conditions.

The possibility of this generalized `defeat device' subversion necessitates continuous monitoring and auditing in real-world conditions, not just simulated conditions at certification time. Such continuous monitoring may benefit from automation, as I discuss in the next section.

\subsection{Compliance: Algorithms Watching Algorithms}

Recently, Amitai and Oren Etzioni proposed a new class of algorithms, called \emph{oversight programs}, whose function is to ``monitor, audit, and hold operational AI programs accountable'' \cite{etzioni2016ai}. Note the emphasis on `operational,' suggesting that these oversight programs are aligned with the point I made earlier about the futility of source code inspection as the only means for regulation.

Oversight algorithms, thus, perform a similar function to today's spam filtering algorithms. But their scope is much wider, as they investigate suspicious behavior by rogue AI algorithms maliciously violating human values. For example, a new class of browser plug-ins is allowing independent, data-driven auditing of the information provided by online advertising platforms to advertisers \cite{auditing}. This has revealed issues in the transparency and accuracy of the current algorithmically-mediated online advertising ecosystem.

One can imagine an algorithm that conducts real-time quantification of the amount of bias caused by a news filtering algorithm--akin to Facebook's recent study \cite{bakshy2015exposure}--and raising an alarm if bias increases beyond a certain threshold.


\subsection{The Limits of Public Engagement}

It is worth highlighting the limits of crowdsourcing of societal values in general, and when it comes to AI in particular. One of the most influential figures in 20th century journalism, Walter Lippman, warned of over-reliance on public opinion when it comes to policy matters that require significant expertise. In Lippman's words,  ``Public opinion is not a rational force.... It does not reason, investigate, invent, persuade, bargain or settle'' \cite{lippmann1927phantom}. This is because it is impossible for a lay person to be fully informed about all facets of every policy question: even an expert practitioner or regulator in one field--say medicine--cannot be sufficiently informed to weigh in on policy matters in another field--say monetary policy. The role of public opinion, Lippman contends, is to check the use of sovereign force, based on assessments made digestible to them by disagreeing experts, pundits and journalists.

There is a lot of merit in Lippman's argument. But he misses a second important role that the public plays: that of shaping moral values and norms. Experts alone cannot dictate what societal values should be. They can influence those values by providing relevant facts, such as the importance of physical exercise in promoting health, or the importance of recycling in the preservation of the environment. But ultimately, norms are shaped through the interaction of various social and evolutionary forces \cite{henrich2004cultural,richerson2005not}. And these values must influence the metrics against which the performance of experts--or AI algorithms--is measured.

\section{Discussion}


The ideas outlined in this article are not entirely new, and many have been discussed in the context of digital democracy \cite{helbing2015society} and the data-driven society \cite{pentland2013data}. Tim O'Reilly recently coined the term \emph{algorithmic regulation} to describe data-driven governance \cite{oreilly2}. To O'Reilly, successful algorithmic regulation must satisfy the following properties (quoted verbatim):
\begin{enumerate}
	\item A deep understanding of the desired outcome
	\item Real-time measurement to determine if that outcome is being achieved
	\item Algorithms (i.e. a set of rules) that make adjustments based on new data
	\item Periodic, deeper analysis of whether the algorithms themselves are correct and performing as expected.
\end{enumerate}
I agree with O'Reilly's characterization. From my perspective, the identification and negotiation of desired outcomes are non-trivial problems. And ensuring that algorithms are performing as expected is not just a technical challenge, but also a social one. This is what makes the social contract framework helpful.

Note that SITL operates at different time-scales than HITL. It looks more like public feedback on regulations and legislations, than feedback on frequent micro-level decisions. Nevertheless, I believe there is value in ensuring we pay attention to all component of `the loop' using an explicit framework. This will be increasingly important as the time between diagnosis and policy adjustment becomes shorter, thanks to progress in data science and machine learning.

I attempted to synthesize various concerns and solutions put forward by many scholars who are thinking about the regulation of algorithmic systems that govern social and economic life. I organized these discussions within two paradigms that have a long history: the human-in-the-loop paradigm from the fields of computer science and supervisory control, and the `social contract' paradigm from political philosophy. The result can be summarized by a call-to-arms that defines the challenge ahead: 

\begin{quote}
\emph{to build institutions and tools that put the society-in-the-loop of algorithmic systems, and allows us to program, debug, and monitor the algorithmic social contract between humans and governance algorithms}. 
\end{quote}

The Age of Enlightenment marked humanity's transition towards the modern social contract, in which political legitimacy no longer emanates from the divine authority of kings, but from the mutual agreement among free citizens to appoint a sovereign. We spent centuries taming Hobbes's Leviathan, the all-powerful sovereign \cite{leviathan}. We must now create and tame the new \emph{Techno-Leviathan}.

\section*{Acknowledgement}


I am grateful for financial support from the Ethics \& Governance of Artificial Intelligence Fund, as well as support from the Siegel Family Endowment.

I am endebted to Joi Ito, Suelette Dreyfus, Cesar Hidalgo, Alex `Sandy' Pentland, Tenzin Priyadarshi and Mark Staples for conversations and comments that helped shape this article. I'm grateful to Brett Scott for allowing me to appropriate the term `Techno-Leviathan' which he originally presented in the  context of Cryptocurrency \cite{technoleviathan}. I thank Deb Roy for introducing me to Walter Lippman's `The Phantom Public' and for constantly challenging my thinking. I thank Danny Hillis for pointing to the co-evolution of technology and societal values. I thank James Guszcza for suggesting the term `algorithm auditors' and for other helpful comments.


\footnotesize


\end{document}